\documentstyle[prb,aps]{revtex} 
\begin{document}
\draft
\twocolumn[\hsize\textwidth\columnwidth\hsize\csname @twocolumnfalse\endcsname
\title{Universality and logarithmic corrections\\
in two-dimensional random Ising ferromagnets}
\author{F. D. A. Aar\~ao Reis, S. L. A. de Queiroz,
and Raimundo R. dos Santos}
\address{
Instituto de F\'\i sica, Universidade Federal Fluminense, Avenida Litor\^anea s/n, 24210-340 Niter\'oi RJ, Brazil}
\date{\today}
\maketitle
\begin{abstract}
We address the question of weak versus strong universality scenarios for
the random-bond Ising model in two dimensions. 
A finite-size scaling theory is proposed, which explicitly incorporates
$\ln L$ corrections ($L$ is the linear finite size of the system) to the
temperature derivative of the correlation length.
The predictions are tested by considering long, finite-width strips of Ising 
spins with randomly distributed ferromagnetic couplings, along which free
energy, spin-spin correlation functions and specific heats are calculated by 
transfer-matrix methods.
The ratio $\gamma/\nu$ is calculated and has the same value as in the 
pure case; consequently conformal invariance predictions remain valid for
this type of disorder.
Semilogarithmic plots of correlation functions against distance yield 
average correlation lengths $\xi^{av}$, whose size dependence agrees very
well with the proposed theory.
We also examine the size dependence of the specific heat, which clearly suggests
a divergency in the thermodynamic limit.
Thus our data consistently favour the Dotsenko-Shalaev picture of logarithmic
corrections (enhancements) to pure system singularities, as opposed
to the weak universality scenario.
\end{abstract}

\narrowtext
\vskip2pc]

\section{Introduction}
\label{intro}

In the study of random magnetic systems, a frequently--asked question is 
whether or not quenched
disorder destroys a sharp phase transition and, in the latter case,
whether critical exponents are the same
as for the corresponding pure magnets~\cite{st83,sh94,sst}.
The Harris criterion\cite{har} provides useful guidance in
a number of cases: if the exponent $\alpha$, characterising the
divergence of the specific heat at the critical point of the
pure system, is positive then randomness induces crossover to a 
different universality class; for negative $\alpha$ the disordered
system is expected to exhibit the same critical behaviour as the
unperturbed one.  However, such a rule is inconclusive
for the subject of this work,
the two-dimensional Ising model, where the specific heat of the pure
system diverges logarithmically (that is, with $\alpha = 0$ )
at the critical point. Further, the Harris approach is perturbative
in the sense that only weak randomness is considered. Non-perturbative
methods are thus required, especially when one wishes to investigate
strongly disordered systems.
A suitable way to deal with this sort
of problems is through numerical calculations on finite systems. One then has
to account for finite-size effects before extrapolating
to the thermodynamic limit. This is done by testing specific
hypotheses bearing upon the nature of asymptotic behaviour.

In the present paper we investigate the theoretical prediction (see Refs.
\onlinecite{sh94,sst} and references therein)
that disorder affects the phase transition of the two-dimensional Ising model
only via a specific, well-defined set of logarithmic corrections to pure-system
critical behaviour; 
here we extend, and give further details of, the results preliminarily reported
in Ref.\ \onlinecite{sbl}.  
Such a prediction is in contrast to recent work~\cite{kim,kuhn,Fahnle92},
according to which critical quantities such
as the zero-field susceptibility and correlation length display power-law
singularities, with the corresponding exponents $\gamma$ and $\nu$ changing continuously with disorder so that the ratio $\gamma/\nu$ is kept
constant  at the pure system's  value (the so-called {\it weak universality} 
scenario\cite{suzuki}). 

Here we calculate free energies and spin-spin correlation functions on
long, finite-width strips of two-dimensional disordered
Ising systems. The main motivation for the use of this geometry is
that strip calculations, together with finite-size scaling (FSS)
concepts~\cite{fisher,fs1} are among the most accurate
techniques to extract critical points and exponents for non-random
low-dimensional systems~\cite{nig82,fs2}.
The rate of decay of correlation functions
determines correlation lengths along the strip.
These latter are, in
turn, an essential piece of Nightingale's phenomenological 
renormalisation scheme~\cite{nig82,fs2}, 
and have been given further relevance via
the connection with critical exponents provided by conformal invariance
concepts~\cite{cardy}.
Early extensions of strip scaling to 
random systems~\cite{early}
have since been pursued further~\cite{sbl,dQ92,sldq} and put into a broader
perspective. In particular, it has been shown that although in-sample fluctuations of correlation functions do not die out as strip length is
increased, averaged values converge satisfactorily\cite{dqrbs}; throughout the
present paper we shall make use of this fact to calculate error bars of
related quantities.

We consider the two-dimensional Ising model on a square lattice
with  bond randomness. The particular version of disorder studied in this
work is a binary distribution of  ferromagnetic
interaction strengths for both vertical and horizontal bonds, 
\begin{equation}
 P(J_{ij})= {1 \over 2} ( \delta (J_{ij} -J_0) +  \delta (J_{ij} -rJ_0) ) \ \ \ ,\ \ 0 \leq r \leq 1 \ \ ,
\label{eq:1}
\end{equation}
\noindent which is the prototypical random-bond Ising system, and exhibits
the unique advantage that its critical temperature $\beta_c = 1/k_B T_c$ 
is exactly known\cite{fisch,kinzel} as a function of $r$ through:
\begin{equation}
\sinh (2\beta_{c} J_{0})\sinh (2\beta_{c}r J_{0}) = 1 \ \ .
\label{eq:2}
\end{equation}
For given $r$ one can then sit at $T = T_c(r)$ and be sure that  numerical errors due to imprecise knowledge of the critical point are absent. Also,
a vast amount of simulational work has been done on this same model (see
Ref. \onlinecite{sst}), thus comparison is made easier when appropriate.

The layout of the paper is as follows. 
We first recall in Sec.\ \ref{logfss} the main predictions~\cite{sh94,sst}
concerning logarithmic corrections to the singular behaviour of
bulk quantities for disordered two-dimensional Ising model, 
and discuss how such corrections should show up in the corresponding 
finite-size quantities. 
In particular, we show that a logarithmic term is expected to be the 
leading correction to the finite-size behaviour of the temperature 
derivative of $\xi^{av}$.
In Sec.\ \ref{calcs} we outline numerical aspects of our calculational 
approach for the magnetic susceptibility, the correlation
length, and the specific heat;
also, the numerical results are presented and discussed.
Sec.\ \ref{conc} summarizes our findings.

\section{Logarithmic corrections and finite-size scaling}
\label{logfss}

For infinite-system quantities close to the critical point, with
$t \equiv (T - T_c)/T_c$, the following forms have been proposed (see Refs.
\onlinecite{sh94,sst} and references therein) for
the  correlation length $\xi_{\infty} (t)$ and initial susceptibility
$\chi_{\infty} (t)$:
\begin{equation}
\xi_{\infty}\sim t^{-\nu}\left[1+C\ln \left(1/t\right)\right]^{\tilde\nu},
\label{xiDS}
\end{equation}
\begin{equation}
\chi_{\infty}\sim t^{-\gamma}\left[1+C\ln \left(1/t\right)\right]^{\tilde\gamma} ,
\label{eq:bulk}
\end{equation}
\noindent where $\nu=1$, $\tilde\nu=1/2$, $\gamma=7/4$, $\tilde\gamma=7/8$
and $C$ is a disorder-dependent positive constant; for $C=0$ one
recovers pure-system behaviour.
Corresponding expressions have been derived for magnetisation and specific heat,
which will not concern us for now. Thus theory predicts that the dominant
power-law singularities (with the same indices as for the uniform system) will
actually be {\it enhanced} by logarithmic divergences. We shall keep to current
use in the field\cite{sh94,sst} and refer to these latter as {\it corrections},
though strictly speaking the term is inappropriate.  

In searching for signatures of such diverging logarithmic corrections  
in systems of finite size, one must be careful about applying recipes used when
the bulk singularity is purely of a power-law nature.     
For instance, a na\"\i ve application, to Eq.\ (\ref{xiDS}), of the  usual shortcut~\cite{fs1} $t\to L^{-1}$ to extract the
size dependence at criticality, would yield a correlation length growing faster
than $L$, which clearly cannot be true. Instead, one must consider the
relationship between bulk quantities predicted by theory and exemplified by
 Eqs.\ (\ref{xiDS}) and (\ref{eq:bulk}), namely
\begin{equation}
\chi_{\infty}\sim (\xi_{\infty})^{\gamma/\nu}\ \ .
\label{eq:chi-xi}
\end{equation}
To see what this implies, recall the FSS hyphotesis~\cite{fisher,fs1} 
for a generic quantity
${\cal Q}_L(t)$ : 
\begin{equation}
{\cal Q}_L(t) = f(L)\ {\cal G}(z)\ \ ,\ \ z \equiv {\xi_{\infty}(t) \over L}\ ,
 \label{eq:fss1}
\end{equation}
where $L$ is the linear lattice size and one assumes small $t$, large $L$.
As is well known, the $L$-dependence must be removed as $z \to 0$.
 It is immediate that, whenever the relationship between 
${\cal Q}_{\infty}(t)$ 
and $\xi_{\infty}(t)$ is a power law such as in Eq.\ (\ref{eq:chi-xi}) above,
$f(L)$ will be a power law as well. This, together with the complementary
condition that only the $L$-dependence must remain asymptotically for $z \gg 1$,
ensures in the case that the finite-size susceptibility at the
critical point must be
\begin{equation}
\chi_L (0) \sim L^{\gamma/\nu} = L^{7/4}\ .
 \label{eq:chifs}
\end{equation}
In other words, FSS implies that
logarithmic corrections must {\it not} show up, and the
finite-size susceptibility at $T_c$
will exhibit pure power-law behaviour against $L$,
with the same power as in the homogeneous case.
The same argument is, of course, valid for $\xi_L(t)$ which must then scale
linearly with $L$ at $T_c$. 
As shown below, numerical data bear out such predictions, for both 
$\chi_L(0)$ and $\xi_L(0)$.
\smallskip\par
This raises the question of how, on finite systems, to seek evidence
for effects of the bulk corrections predicted 
in Eqs.\ (\ref{xiDS})--(\ref{eq:bulk}).
In the following we show that the proper quantities to consider are
temperature derivatives of e.g. $\xi_L(t)$. We apply standard
FSS concepts to show that, although the dominant behaviour of such quantities
is in powers of $L$, the leading {\it corrections} to FSS must depend on
$\ln L$. This is in contrast with the corresponding (non-diverging)
corrections to FSS for,
say, $\xi$ which can be fitted by inverse power laws (see 
Ref.\onlinecite{sldq} and below). 

First we recall that the FSS form for $\xi$ is, from Eq.\ (\ref{eq:fss1}):
\begin{equation}
\xi_L(t) = L\ \phi(z)\ ,\ \ \ \phi(z) \to \cases{z,\quad\qquad z \ll 1 \cr
                                                 const.,\ \ \ z \gg 1} 
\label{eq:xifs}
\end{equation}

The temperature derivative of $\xi_L$ is then
\begin{equation}
\mu_L(t) \equiv {d\xi_L(t)\over dt} = \mu_{\infty}(t) \phi^{\prime}(z)\ ,\ \ \
\phi^{\prime}(z) \to \cases{1, \ \ z \ll 1 \cr
                           0, \ \  z \gg 1} 
\label{eq:mul1}
\end{equation}
where $\mu_{\infty}(t) \equiv
d\xi_{\infty}/dt$ and the prime denotes a
 derivative with respect to $z$. While the bulk
limit $z \ll 1$ of Eq.\ (\ref{eq:mul1}) is a straightforward identity,
the vanishing of 
$\phi^{\prime}(z)$ for $ z \gg 1$ [as implied by Eq.\ (\ref{eq:xifs})]
must be qualified. Indeed, $\mu_L(t)$ does not diverge in the latter
regime, while $\mu_{\infty}(t)$ does when $t \to 0$. Thus
$\phi^{\prime}(z) \sim \left(\mu_{\infty}(t)\right)^{-1}$, in the sense that
the dependence of $\phi^{\prime}$ on $t$ through $\xi_{\infty}$
must be such as to cancel the diverging $t$--dependence of $\mu_{\infty}$.
Since the FSS {\it ansatz} predicts that $t$
only arises through
the ratio  $\xi_{\infty}(t)/L$, one can deduce the $L$--dependence of
$\mu_L(t)$ for $z \gg 1$. Up to now, the argument is entirely general and variations of it
have been commonly used in the FSS literature.

Turning to the two-dimensional random-bond Ising model,
where the bulk quantities are expected to behave as in 
Eqs.\ (\ref{xiDS}--\ref{eq:bulk}), one has for $t \ll 1$ (consistent with our goal of
deriving expressions suitable for the $z \gg 1$ regime)~:
\begin{equation}
\xi_{\infty}\sim t^{-\nu}\left(\ln 1/t\right)^{\tilde\nu},
\label{eq:xiapp}
\end{equation}
which can be iteratively inverted to give $t$ as a function of $\xi_{\infty}$:
\begin{equation}
t \sim \xi_{\infty}^{-1/\nu}\left(\ln\xi_{\infty}\right)^{\tilde\nu/\nu} .
\label{eq:tapp}
\end{equation}
The expression for $\mu_{\infty}(t)$ is
\begin{equation}
\mu_{\infty}(t)\sim
t^{-(1+\nu)}\left[1+C\ln \left(1/t\right)\right]^{\tilde\nu},\\
\label{eq:bulkp}
\end{equation}
plus less-divergent terms, which for $t \ll 1$ can be put as 
\begin{eqnarray}
\mu_{\infty}(t)&\sim&
t^{-(1+\nu)}\left(\ln 1/t\right)^{\tilde\nu}
\nonumber\\
&&\simeq \xi_{\infty}/t =
\xi_{\infty}^{1+1/\nu}\left(\ln\xi_{\infty}\right)^{-\tilde\nu/\nu} \ 
\label{eq:muapp}
\end{eqnarray}
where Eq.\ (\ref{eq:tapp}) was used in the last step. It follows immediately
that
\begin{equation}
\phi^{\prime}(z) \sim z^{-(1+1/\nu)} (\ln z)^{\tilde\nu/\nu}\  
\label{eq:phip}
\end{equation}
which, when plugged back into Eq.\ (\ref{eq:mul1}) together with 
Eq.\ (\ref{eq:muapp}), gives:
\begin{equation}
\mu_L \sim L^{1+1/\nu} \left[1 - \ln L /\ln \xi_{\infty}\right]^{\tilde\nu/\nu} 
,\ \ \ \ z \gg 1 ,
\label{eq:mul2}
\end{equation}
so all diverging factors related to $\xi_{\infty}$ are removed, but a
non-diverging $\xi_{\infty}$-dependent term remains which eventually vanishes. 
Strictly speaking, Eq.\ (\ref{eq:mul2}) means that for  both
$t \ll 1$ {\it and} $L \gg 1$, but such that $z \gg 1$, one must observe essentially
the leading power-law form $ \mu_L \sim L^{1+1/\nu}$. However, even though
Eq.\ (\ref{eq:2}) enables one to sit exactly at $t=0$, Eq.\ (\ref{eq:mul2})
suggests the existence of a regime in which the leading correction to
power-law behaviour is $\sim (1 - A\ln L)^{\tilde\nu/\nu}$
for finite and not very large strip
widths $L$, thus defining an effective (non-diverging) screening
length $\xi_s \equiv e^{1/A}$. 
This heuristic procedure draws on ideas used to interpret experimental data
for systems where a full divergence of the correlation length is hindered by percolation\cite{perc}, random field\cite{rand} or frustration\cite{frust}
effects. Defining  the inverse correlation lengths $\kappa$ (actually observed), $\kappa_0 \sim t^{\nu}$  and $\kappa_s \equiv (\xi_s)^{-1}$ (representing the
physical factor that smears the divergence, e.g. domain size), one writes
\begin{equation}
\kappa = \kappa_0 + \kappa_s\ \,  
\label{eq:kappa}
\end{equation}
with good results\cite{perc,rand,frust}. Here, $\kappa_s$ does not originate
from a physical feature of the infinite system; instead, it reflects  
the overall effect of higher-order
corrections in such pre-asymptotic region (strip widths $L \lesssim 15$). 
While $\xi_s$ is of a different nature to the crossover length $L_C$ setting 
the scale above which disorder effects are 
felt~\cite{sh94,sst}, the two lengths vary similarly with disorder,
as explained below.

We now describe the numerical procedures used to test the predictions
given by Eqs.\ (\ref{eq:chifs}) and (\ref{eq:mul2}), and the respective results.

\section{Calculational method and results}
\label{calcs}
 
We have used long strips of a square lattice, of width $4 \leq L \leq 14$
sites with periodic boundary conditions. 
In order to provide samples that are sufficiently representative of disorder, 
we iterated the transfer matrix\cite{fs2} typically along $10^7$
lattice spacings.

At each step, the respective vertical and horizontal bonds between first-neighbour spins $i$ and $j$ were drawn from the probability distribution
Eq.\ (\ref{eq:1}) above. 
We have mainly used three values of $r$ in calculations: $r=0.5$, $0.25$ and
$0.1$; the two smallest values have been chosen for the purpose of
comparing with recent Monte-Carlo simulations where $\nu$ and
$\gamma$ are evaluated~\cite{kimunp}. 
The critical temperatures, from Eq.\ (\ref{eq:2}), are: 
$T_c\, (0.5)/J_0 = 1.641 \dots$~; 
$T_c\, (0.25)/J_0 = 1.239 \dots$ ; 
$T_c\, (0.1)/J_0 =  0.9059 \dots$ 
(to be compared with  $T_c\, (1)/J_0 = 2.269 \dots$ ). We also evaluated
critical correlation lengths and their derivatives for $r=0.01$ and $0.001$, 
with respective critical temperatures $T_c/J_0 = 0.5089\dots$ and $0.3426
\dots$ .
 
\subsection{Susceptibility}
\label{su}

The calculation of finite-size susceptibility data and their
extrapolation goes as follows. 
First, we include a uniform longitudinal field $h$ in the Hamiltonian,
and obtain the largest Lyapunov exponent $\Lambda_{L}^{0}$ for a strip of width $L$ and length  $N \gg 1$ in the usual way\cite{early,ranmat}.
Starting from an arbitrary
initial vector ${\bf v}_0$, one generates the transfer matrices ${\cal T}_{i}$
that connect columns $i$ and $i+1$, drawing bonds from the distribution 
Eq.\ (\ref{eq:1}), and applies them successively, to obtain:
\begin{equation}
 \Lambda_{L}^{0} = {1 \over N} \ln \Biggl\{ {\Bigl\Vert \prod_{i=1}^N {\cal T}_{i} {\bf v}_0 \Bigr\Vert \over \bigl\Vert {\bf v}_0 \bigr\Vert }\Biggr\}\ \ .
\label{eq:lyap}
\end{equation}
The average free energy per site is then 
$f_{L}^{\ ave}(T,h) = - {1 \over L} \Lambda_{L}^{0}$, in units of $k_{B}T$.
The initial susceptibility of a strip, $\chi_L (T_c)$, is given by
\begin{equation}
\chi_L (T_c) ={ \partial^{2}  f_{L}^{\ ave}(T,h) \over \partial
h^2}\Biggr|_{T=T_c,h=0}
= L^{\gamma/\nu}\ Q(0)\ \ ,
\label{eq:5}
\end{equation}
\noindent where, according to the
discussion in the preceding Section, we assume a pure power-law dependence
on $L$ at $T = T_c$.

 As $f_{L}^{\ ave}(T,h)$ is
expected to have a normal distribution~\cite{derrida,ranmat}, so will
$\chi_L$. Thus the fluctuations are  Gaussian, and relative errors 
must die down with sample size (strip length) $N$ as $1/\sqrt{N}$. 
Typical strip lengths varied from $N=2\times10^6$ (for $r=0.5$) to 
$N=2\times10^7$ (for $r=0.1$), which are much longer than those
used in Ref.\onlinecite{sldq}; they provide estimates for the
free energy with an accuracy of 0.01\%, which is crucial to
compute reliable numerical derivatives.
In order to get rid of start-up 
effects, the first $N_0 = 10^5$ iterations were discarded. 
The intervals (of external field values, in this case) used in  
obtaining finite differences for the calculation of numerical derivatives
must be strictly controlled, so as not to be
an important additional source of errors. We have managed to minimise
these latter effects by using $\delta h$ typically of order $10^{-4}$
in units of $J$ when calculating  $f_{L}^{\ ave}(T_c; h=0,\, \pm \delta h)$
for the derivative in Eq.\ (\ref{eq:5}). We estimated the first Lyapunov
exponent at $(T=T_c,h=0)$ and $(T=T_c,h=\pm\delta h)$ with four different
realizations of the impurity distribution, each one giving a separate
estimate of the initial susceptibility. From them the average $\chi_L (T_c)$
and the error bars (twice the standard deviation among the four overall
averages) are taken.

A succession of estimates, $\left(\gamma/\nu\right)_L$, for the 
ratio $\gamma/\nu$ is then obtained from Eq.\ (\ref{eq:5}) as follows:
\begin{equation}
\left({\gamma\over\nu}\right)_L=
{\ln \left[\chi_L(T_c)/\chi_{L-1}(T_c)\right]
\over
\ln \left[L/(L-1)\right]}\\
\label{eq:5p}
\end{equation}
The respective error bars follow from those of the
corresponding finite-size susceptibilities.
In order to extrapolate this sequence, we refer to early work on the eigenvalue
spectrum of the transfer matrix for pure systems with a marginal
operator in the Hamiltonian~\cite{car86}. There, it is shown that the critical
free energy per site is affected only by an additive logarithmic 
term in the coefficient of the leading, $L^{-2}$--dependent, finite-size
correction (proportional to the conformal anomaly\cite{bcna}, $c$):
 $f(L) - f(\infty) = -(\pi/6L^2)[c +B(\ln
L)^{-3} + \dots]$. Since disorder is expected to be marginally irrelevant
in the present case, and assuming that the field derivatives commute with 
the $L$--dependence (at least as dominant terms are concerned), we expect a
similar picture to hold here. Of course, with the imprecisions introduced by
randomness one can only expect to see the leading power-law dependence
(see, e.g., Ref.~\onlinecite{sldq} for further illustrations of this point).
   
Least-squares fits for plots of $\left(\gamma/\nu\right)_L$ against
$1/L^2$ provide the 
following extrapolations: $\gamma/\nu=1.748\pm 0.012,\ 1.749\pm 0.008,$ and
$1.746\pm 0.013$, respectively for $r=0.50,\ 0.25,$ and 0.10; 
the latter two estimates agree with 
$1.74\pm 0.03,\ 1.73\pm 0.05$, obtained in Ref.\onlinecite{kimunp}.

The overall picture is thus consistent with the prediction of
Eq.\ (\ref{eq:chifs}), that is $\gamma/\nu = 7/4$, same as for the
pure system, for all degrees of disorder. Recalling the Introduction,
this still is not enough to distinguish between weak- and
strong-universality scenarios, as both coincide in their predictions for the
{\it ratio} of exponents. One needs to try and isolate a single exponent,   which will be done in the next subsection through investigation of correlation lengths.

Taken together with the results of Ref.~\onlinecite{sldq} where $\eta$ 
was found to 
be $1/4$ through an analysis of averaged correlation lengths, and using the 
scaling relation $\gamma/\nu = 2 - \eta$, the present analysis of
finite-size susceptibilities gives independent support to the view that: 
(1) the conformal invariance relation~\cite{cardy} 
$\eta =L/\pi \xi_{L}(T_c)$
still holds for disordered systems, provided that an 
{\it averaged} 
-- as opposed to {\it typical}, see next subsection -- 
correlation length is used; 
and that (2) the appropriate correlation length to be used is that coming 
from the slope of semi-log plots of correlation functions against
distance~\cite{sldq}.
Interestingly, the connection with the conformal invariance prediction
also rules out any explicit diverging logarithmic $L$-dependence on $\xi_L$.

\subsection{Correlation lengths}
\label{cor}

The aim of this subsection is to check on the validity of 
Eq.\ (\ref{eq:mul2}),
or rather, its predicted consequences in the pre-asymptotic region 
within our reach, $t \ll 1$, $L \lesssim 15$.

The first difference to the
free energy calculation described above is that the correlation functions are
expected to have a distribution close to {\it log}-normal~\cite{dh,derrida} rather than a normal one. This has been thoroughly checked recently\cite{dqrbs}.
Thus self-averaging is not present, and fluctuations for
a given sample do $not$ die down with increasing sample size. However, it has been numerically verified that the spread among overall averages 
({\it i.e.} central estimates) from different
samples does shrink (approximately as $N^{-1/2}$) as the samples' size 
($N$) increases (see Fig. 2 of Ref. \onlinecite{dqrbs}). 
Accordingly, in what follows the error bars quoted arise from fluctuations
among four central estimates, each obtained from a different impurity
distribution. Similar procedures seem to have been followed in
Monte-Carlo calculations of correlation functions in finite ($ L \times L$)
systems~\cite{talapov}.

The direct calculation of correlation functions, 
$\langle\sigma_{0} \sigma_{R}\rangle$, 
follows the lines of Section 1.4 of Ref. \onlinecite{fs2}, with standard 
adaptations for an inhomogeneous system\cite{sldq}. 
For fixed distances up to $R=100$, and for strips with the same length
as those used for averaging the free energy, the correlation functions
are averaged over an ensemble of $10^4$--$10^5$ different estimates 
to yield $\overline{\langle\sigma_{0} \sigma_{R}\rangle}$. 

The average correlation length, $\xi^{av}$, is defined by
\begin{equation}
\overline{\langle\sigma_{0} \sigma_{R}\rangle}
\sim
\exp\left(-R/\xi^{av}\right),\\
\label{eq:xi}
\end{equation}
and is calculated from least-squares fits of straight lines to semi-log 
plots of the average correlation function as a function of distance,
in the range $10\leq R\leq 100$. 
And, finally, $\xi^{av}$ is in turn
averaged over the different realizations of impurity distributions.

Recall that, as explained in Ref. \onlinecite{sldq}, 
the inverse of $\xi^{av}$ is not the same 
as the difference between the two leading
Lyapunov exponents, which gives the decay of the {\it most probable}, or
typical
(as opposed to averaged) correlation function\cite{sldq,ranmat,crisanti}.
It has been predicted\cite{ludwig} that typical correlations in bulk two-dimensional random Ising magnets decay as 
$ \langle \sigma_0 \sigma_R \rangle \sim R^{-1/4} ( \ln R)^{-1/8}$, while
for averaged ones as in Eq.\ (\ref{eq:xi}) logarithmic corrections are
washed away, resulting in a simple power-law dependence.
 For strips one could expect, in analogy with
the case of pure systems with marginal operators\cite{car86}, additive
logarithmic corrections to the leading $L^{-1}$ behaviour of typical
correlations: 
$\Lambda_{L}^{1} - \Lambda_{L}^{0} = (\pi/L)[\eta +D(\ln L)^{-1} + \dots]$
with $\eta =1/4$. 

It has been conjectured that the averaged correlation functions at criticality of the random-bond Ising model are identical to those
of the pure case\cite{talapov}; numerically the two quantities are indeed very
close\cite{dqrbs,talapov}, while most-probable and pure-system correlation
functions do no fit each other so well, though their $L$--dependence is
similar\cite{dqrbs}. 
Given the exact result\cite{dds} that, for strips of pure Ising spins the 
corrections to the leading $L^{-1}$ behaviour of $(\xi^{av})^{-1}$ as given
by Eq.\ (\ref{eq:xi}) depend on $L^{-2}$, it seems reasonable to expect $L^{-x}$
({\it i.e.} faster than inverse logarithmic) terms also in the present case.
This has been shown to work well,
with the same $x=2$, in Ref.\onlinecite{sldq}.

We now proceed to testing Eq.\ (\ref{eq:mul2}). 
We calculate $\mu_L$ at $T_c$ [see Eq.\ (\ref{eq:mul1})] numerically, from 
values of $\xi^{av}_L$ evaluated at $T_c\pm\delta T$, with $\delta T/T_c=10^{-3}$.
  
Assuming a simple power-law divergence $\xi_{\infty} \sim t^{-\nu}$, -- i.e., ignoring, for the time being,
less-divergent terms such as logarithmic corrections --  
we obtain the estimates for systems of sizes $L$ 
and $L-1$ :
\begin{equation}
{1 \over \nu_L} = 
{ \ln \left(\mu_L/\mu_{L-1}\right)_{T=T_c}
\over \ln (L/L-1)} - 1\ .
\label{eq:7}
\end{equation}

This is slightly different from the usual fixed-point 
calculation~\cite{fs2}, and is more convenient in 
the present case where the exact critical temperature is known. 
Our data for each pair of ($L, L-1$) strips have appeared in 
Ref.\ \onlinecite{sbl}, and we quote here, for completeness,
just the extrapolated (against $1/L^2$) values: 
$\nu=1.032\pm 0.031$ (for $r=0.5$; here
we have extended the previous calculations up to $L=14$),
$\nu=1.083\pm0.014$ ($r=0.25$), and $\nu=1.14\pm0.06$ ($r=0.10$). 
Taken at face value, these data show a systematic trend towards values 
of $\nu$ slightly larger than the pure-system value of 1, though the 
variation is smaller than that shown in Ref.~\onlinecite{kimunp}. 
\smallskip\par
Before accepting this trend as an indication of the weak-universality
scenario, we must test for corrections to pure-system behaviour caused by
less-divergent terms, as being responsible for the apparent change of
$\nu$ with disorder.
We then try to check whether our data fit a form inspired by Eq.\ (\ref{eq:mul2}) with $\nu =1$ (the pure-system value) and $\tilde\nu =1/2$, namely
\begin{equation}
 {\mu_L\over L^2}\sim \left(1 - A \ln L\right)^{1/2}\ ,\\
\label{eq:mul3}
\end{equation}
Prior to displaying our results, we recall that the influence of randomness
is expected to show 
on scales larger than a disorder-dependent characteristic length 
$L_C$~\cite{sh94,sst}. For $L < L_C$ one should have apparent pure-system
behaviour.

 A plot of $\left(\mu_L/L^2\right)^2$ as a function of $\ln L$ for different values of $r$, including $r = 1$, is shown in 
 Figure \ref{figrb}. The pure-system behaviour consists in a monotonic
approach to a horizontal line, with ever-decreasing slope.
For $r=0.50$ and $0.25$ we
can see the pure-system trend for small $L$, followed by a clearly marked
crossover towards a form consistent with Eq.\ (\ref{eq:mul3}).
In each case, log-corrected behaviour sets in for suitably large $L$,
exactly in the manner predicted by theory: the data stabilize 
onto a straight line with negative slope only for $L\gtrsim L_C$,
which decreases with increasing disorder\cite{sh94,sst}. One may assume, admittedly with some arbitrariness, $L_C$ for each $r$ to be approximately
the location of the maximum of the respective curve in  
Figure \ref{figrb}. This gives $L_C \simeq 8$, 5, and 2 respectively
for $r=0.50$, $0.25$ and $0.10$ (for $r=0.10$
data for $L=2$ and $3$, not shown in the Figure, were used as well).

An order-of-magnitude guide to the size of the pre-asymptotic region
where Eq.\ (\ref{eq:mul3}) is expected to hold, such that for larger $L$
the pure power-law behaviour predicted by Eq.\ (\ref{eq:mul2}) at $t=0$
takes over, is the ``screening length'' $\xi_s \equiv e^{1/A}$ of
Eq.\ (\ref{eq:mul3}). For $r=0.50$, $0.25$ and $0.10$ one has the 
approximate values $\xi_s \sim 4\times 10^{16}$, $7\times 10^4$ and 
$4\times 10^2$ respectively.
Though any of these
is far beyond the largest strip width within reach of calculations, the trend
against disorder is clearly similar to that of $L_C$.  

We thus tried stronger disorder (smaller values of $r$), in order to look for  
a signature of pure power-law behaviour at a feasible $L \lesssim 15$.
In Figure \ref{figrb2} we show $\left(\mu_L/L^2\right)^2$ as a function of
$\ln L$ for $r=0.01$ and $0.001$. Proximity to the percolation threshold is
reflected in the large
error bars, which render our central estimates virtually meaningless
for $L \lesssim 5$;
for larger $L$, fluctuations are reduced, owing to the exponential growth 
in the number of
intra-column configurations, so we still can manage reasonable fits in that
range. Unfortunately, no clear sign can be seen of a trend towards a horizontal
line. We believe that a conjunction of $(i)$ smaller $r$, $(ii)$ larger $L$ 
and $(iii)$ longer strip length $N$ would eventually unearth the expected
stabilization, though we do not feel secure to venture numerical guesses
at this point.     

The above correlation length analysis thus provides us with an  
interpretation of the numerical data which, it should be stressed, is 
backed by theory\cite{sh94,sst}, without resorting to disorder-dependent
exponents. 
Nevertheless, we have found that the general statistical quality of the 
data does not 
allow one to distinguish clearly in favour of either possibility, in terms,
e.g., of least-squares fits. We therefore seek complementary quantitative 
information through the analysis of specific heat data.

\subsection{Specific heats}
\label{sph}

The same theory\cite{sh94,sst} that gives rise to Eqs.\ (\ref{xiDS}-\ref{eq:bulk})
predicts that the singular part of the bulk specific heat per particle for
the disordered Ising
model, near the critical point, is given by
\begin{equation}
C_{\infty}(t) \simeq (1/ C_0) \ln \left( 1 + C_0 \ln (1/t)\right)\ \ ,
\label{eq:bsph}
\end{equation}
where again $C_0$ is proportional to the strength of disorder, and the pure-system simple logarithmic divergence is recovered as 
$C_0 \to 0$. For $C_0 \neq 0$ and $t \ll 1$ a double-logarithmic singularity
arises, whose amplitude  Eq.\ (\ref{eq:bsph}) predicts to 
decrease as disorder increases. 
The bulk specific heat cannot then be put as a simple function of the
correlation length given
in Eq.\ (\ref{xiDS}), and one cannot predict pure-system behaviour
against $L$ for finite systems, as was the case for the susceptibility
and correlation length above. Instead, theory gives\cite{sst}
\begin{equation}
C_L(t=0) \simeq C_1 + a \ln \left(1 +b \ln L\right)\ \ ,
\label{eq:fsssph}
\end{equation}
where, similarly to Eq.\ (\ref{eq:bsph}), $b \to 0$ for vanishing disorder.
In this latter limit the product $ab$ must remain finite,
but it is not {\it a priori} obvious from theory
whether $a \propto b^{-1}$ away from that. 
In fact, the form Eq.\ (\ref{eq:fsssph}) has been verified by Monte-Carlo
simulations of $L \times L$ systems\cite{wang} with the result that the
slope of plots of $C_L$ against $\ln \ln L$ {\it decreases} for increasing
disorder. This shows that in this case the simple FSS recipe $t \to L^{-1}$
seems to work satisfactorily. 

An investigation of the specific heat on strips is clearly of interest, 
in order to check the consistency of our own correlation-length data,
and also to provide a comparison with the trends found for the specific heat in
$L \times L$ systems, both as described above and in recent work\cite{kim}
where a non-diverging behaviour is apparently found in the thermodynamic limit.

Our results are displayed in Figure \ref{figsph}, where one can see that the
fit to a double-logarithmic form is reasonable; for small disorder $r=0.5$
we get an overall better fit to a pure logarithmic divergence, 
similarly to the result
for  $L \times L$  lattices\cite{wang}.
This is again because, as disorder decreases one gets apparent pure-system
behaviour for relatively large $L$.

The slope of the plots turns smaller for higher disorder, again in agreement
with the trend found for  $L \times L$  lattices\cite{wang}; however,
no sign of an eventual trend towards non-divergence\cite{kim} can be distinguished. 

The recent claims that for strongly disordered Ising systems in two dimensions,
the specific heat is finite at $T_c$, have been made on the basis of numerical
simulations of site-diluted models\cite{kim}.
Specific heats were plotted against $t$ ($t > 0$) for system sizes
and temperatures such that $L/\xi_{\infty}(t) > 1$
(thus excluding the very close vicinity of the transition); see Fig. 1 of 
Ref.~\onlinecite{kim}.
While for impurity concentration $c=1/9$ a divergence
was clearly seen, data for $c=1/4$ and $1/3$ were interpreted
as signalling a finite bulk specific heat at the transition.
Such findings have been criticised\cite{comments}.
At this point it is worth recalling experimental data.
First, in bulk systems the specific heat exhibits
a broad regular background against which the singular part
must be singled out. Early experiments on the two-dimensional site-diluted
Ising system Rb$_2$Co$_x$Mg$_{1-x}$F$_4$
showed that the amplitude of the singular part
of the magnetic specific heat {\it decreases} as dilution
$1-x$ increases\cite{ikeda}. However, owing to experimental difficulties,
chief among them the smearing of $T_c$ due to sample inhomogeneities,
clear peaks could be found only for $1-x \lesssim 0.11$. 
Later, results from the more accurate technique of birefringence\cite{biref1}
and with a presumably higher-quality sample confirmed\cite{biref2} that 
the specific heat diverges for $1-x = 0.15$, apparently with a single
logarithmic dependence identical to that of the pure system; this was ascribed
to an extreme narrowness of the region  where disordered (double-logarithmic)
behaviour would show up (in agreeement with theory\cite{sst,biref2}). 
Though, to our knowledge, a systematic study
of the variation of specific heat of two-dimensional Ising systems against
dilution, by e.g. birefringence techniques, has not been done,
useful hints may be taken from the
corresponding three-dimensional case of Fe$_x$Zn$_{1-x}$F$_2$. There, 
birefringence experiments\cite{3d} show that as dilution increases, the 
relative position of
the (narrow) peak at $T_c$ against that of the (broad) maximum of a
short-range order background contribution switches from higher to lower
temperatures. This fact is not directly related to the particular
three-dimensional features which are used to explain the dilution dependence
of the specific heat amplitude\cite{3d}.
Thus, it is not unlikely that for two dimensions too the apparent non-diverging
behaviour seen, for $T > T_c$  at $c=1/4$ and $1/3$, represents only the background. To see the actual (probably small) peak one would have to go closer
to $T_c$; imprecisions in the knowledge of $T_c$ for site-diluted systems
(see e.g. Ref. \onlinecite{dQ92}) may be of capital importance then.

In contrast to this, here and in Ref. \onlinecite{wang} one sits right $at$
the exactly known $T_c$. Further, according to the discussion of finite-size
specific heats above, the amplitude of the peak at the bulk transition
translates directly into the slope of the plot of $C_L$ against $\ln \ln L$, 
so the regular background is easily dealt with.       
  
In short, the evidence presented here clearly indicates that the specific heat
diverges at the transition, with a double-logarithmic behaviour. Thus the
critical exponent $\alpha$ is non-negative. Through hyperscaling arguments,
this ties in with our findings for the correlation length,
as shown in the following. 
For weak enough disorder, there should be no question about the dimensionality
of the system, as opposed to near the percolation threshold in the 
{\it diluted} case [corresponding to $r=0$ for the bond distribution
Eq.\ (\ref{eq:1})], where one might argue in favour of substituting the 
{\it fractal} dimension for the actual lattice dimensionality. 
Therefore, hyperscaling should be fully applicable with $d=2$, which yields
\begin{equation}
{\alpha\over 2}=1-\nu.
\label{hyper}
\end{equation}
Since our specific heat data implies $\alpha\geq 0$ (most likely $\alpha = 0$),
one must have $\nu\leq 1$, thus excluding the disorder-varying exponents
given in Sec.\ IIIb and in previous works.\cite{kim,kuhn,Fahnle92,kimunp}

\section{Conclusions}
\label{conc}

We have addressed the question of strong versus weak universality in the 
two-dimensional random-bond 
(i.e., exchange couplings being either $J$ or $rJ$ with equal probability; 
$0<r\leq1$ measures the degree of disorder) 
Ising model, through extensive transfer matrix calculations.
A key ingredient in the analysis of our data has been the consideration of
subtle finite-size scaling (FSS) effects; 
these come about as a result of constraints imposed by the Dotsenko-Shalaev
theory\cite{sh94} for logarithmic corrections in the thermodynamic limit. 
We have established that while the correlation length (and the susceptibility) 
itself should display no signature of size-dependent logarithmic corrections,
its temperature derivative, $\mu_L\equiv d\xi_L/dt$, shows a $\ln L$ dependence 
($L$ is the strip width) over a wide range of system sizes.
Actually, at the (exactly known) critical temperature for the infinite system
and for constant disorder (i.e., fixed $r$) the behaviour with linear
size is as follows. 
For $L<L_C$, with $L_C$ being a crossover length, the system behaves as in the
pure case; $L_C$ decreases monotonically with disorder and $2 \le L_C \le 8$ 
for the values of $r$ we considered.
Above $L_C$, $\mu_L$ is dominated by a $\ln L$ enhancement {\it over the usual 
pure system power law}; 
that is, the numerical data can be explained through consistent theories, 
without resorting to disorder-varying critical exponents. 
The FSS theory developed here also suggests that as $L$ increases,
beyond a (heuristically introduced) screening length $\xi_s$,
one will eventually reach an asymptotic regime where the logarithmic
enhancements will vanish, leaving only pure power-law (pure-system-like)
behaviour; see Sec.\ \ref{logfss}.
This coherence length tracks the crossover length, in the sense of 
decreasing with increasing disorder, but its order of magnitude is
way beyond the reach of our numerical capabilities ($\xi_s\gtrsim 10^2$)
for us to venture a more refined analysis of this issue.
Note, however, that when $t\to 0$ {\it after} the thermodynamic limit
has been taken (which is an entirely different matter) it is 
expected that $\ln (1/t)$ corrections, as predicted by the 
Dotsenko-Shalaev theory, should manifest themselves.
Also, our data independently confirm that the conformal invariance
result $\xi^{av}=L/\pi\eta$ is still valid for the two-dimensional random--bond
Ising model, with $\eta=1/4$ as in the pure case.

As a further test on the consistency of the proposed scenario, we have
examined the size-dependence of the specific heat for this system. 
Consistently with the above findings, the specific heat was seen
to be clearly divergent in the thermodynamic limit. 
Since there are no physical grounds to invoke a mechanism leading to 
changes in the hyperscaling relation, the case for weak universality
cannot be supported by our data. 
Further, it must be noted that a variety of studies of this problem, 
both theoretical\cite{dss,heuer} and experimental\cite{ikeda,biref2,hagen}
concurs with the idea that the leading singularities remain the same as 
in the pure case, though they have not dealt with the detection of 
logarithmic corrections.

As regards works whose conclusion is that weak-universality holds
instead\cite{kim,kuhn,kimunp}, though $\xi_L(T)$ and the susceptibility 
$\chi_L(T)$ were calculated, no attempt seems to have been
made to fit the corresponding data to a form similar to Eq.\ (\ref{eq:mul3}).
Thus it remains to be checked whether they would also be consistent with
suitable FSS expressions based on strong universality concepts. 
 
\acknowledgements 
We thank Laborat\'orio Nacional de Computa\c c\~ao Cien\-t\'\i\-fica 
(LNCC) for use of their computational facilities, and
Brazilian agencies  CNPq
and FINEP, for financial support. Special thanks are due to R. B.
Stinchcombe for invaluable discussions, and to D. Stauffer for useful
suggestions.

\begin{figure}
\caption{Finite-size scaling plots of logarithmic corrections 
[Eq.\ (\protect{\ref{eq:mul3}})].
Straight lines are least-squares fits of data respectively for
$L= 9 - 14$ ($r=0.5$); $7 - 12$ ($r=0.25$)  and  $4 - 12$ ($r=0.1$).
The error bars are smaller than the data points.}
\label{figrb}
\end{figure}

\begin{figure}
\caption{Finite-size scaling plots of logarithmic corrections 
[Eq.\ (\protect{\ref{eq:mul3}})] for strong disorder.
Straight lines are least-squares fits of data respectively for
$L= 6 - 12$ ($r=0.001$) and $7 - 12$ ($r=0.01$) .} 
\label{figrb2}
\end{figure}

\begin{figure}
\caption{Specific heat per site at criticality for $L= 4 - 12$ and $r=0.50$ (squares), $0.25$ (crosses) and $0.1$ (triangles),
against $\ln \ln L$ [Eq.\ (\protect{\ref{eq:fsssph}})].}
\label{figsph}
\end{figure}

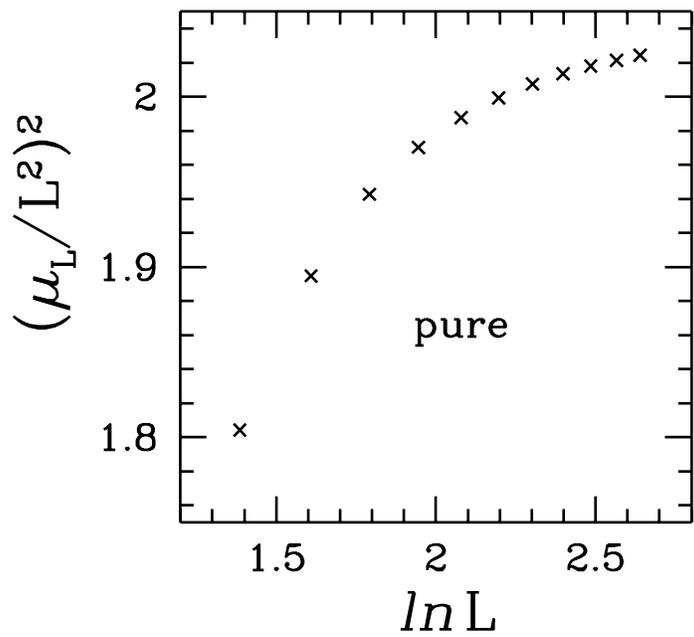
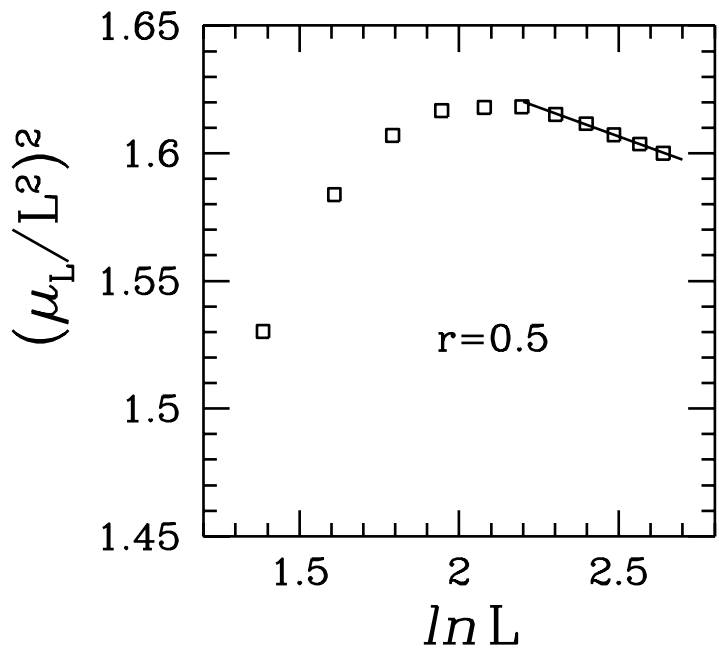
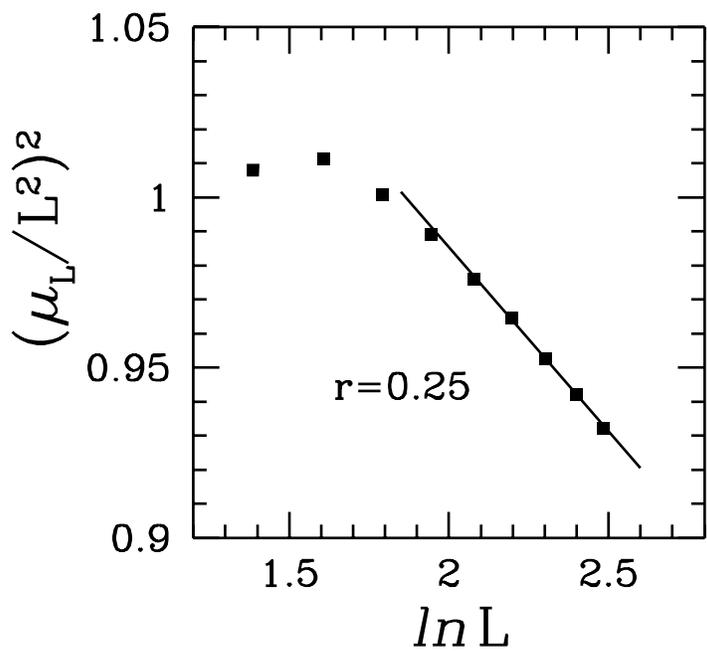
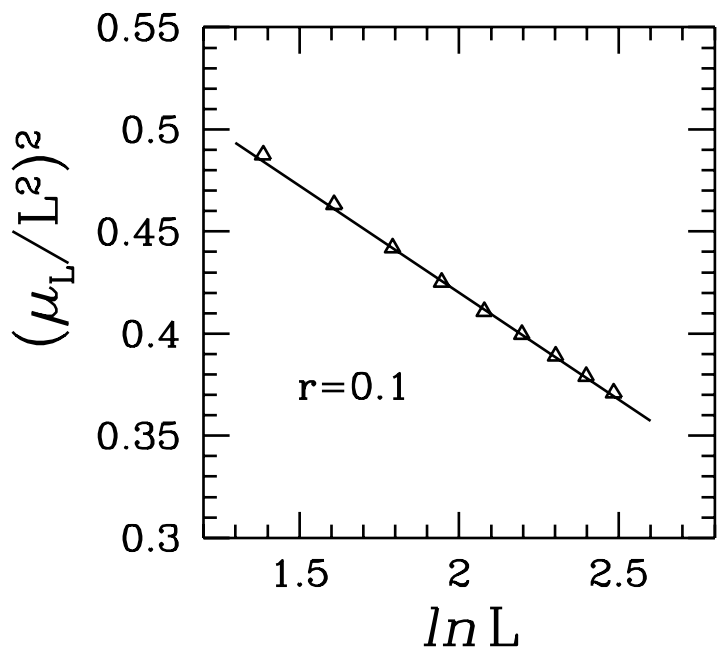
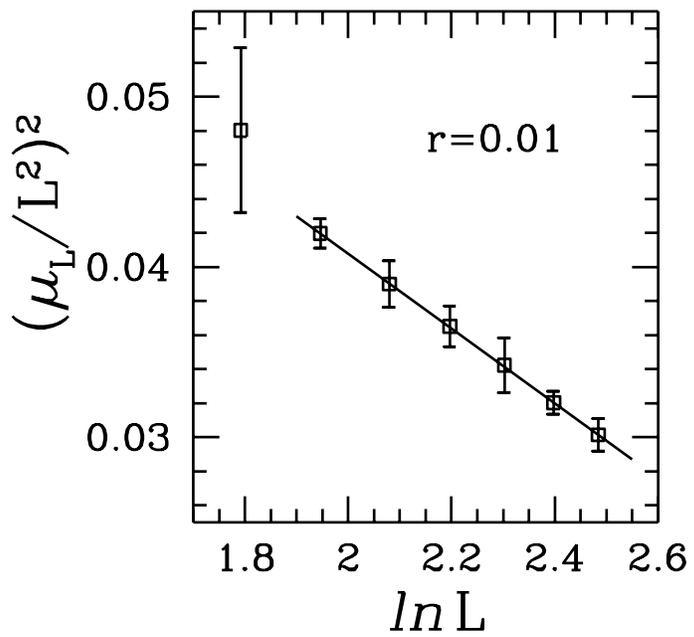
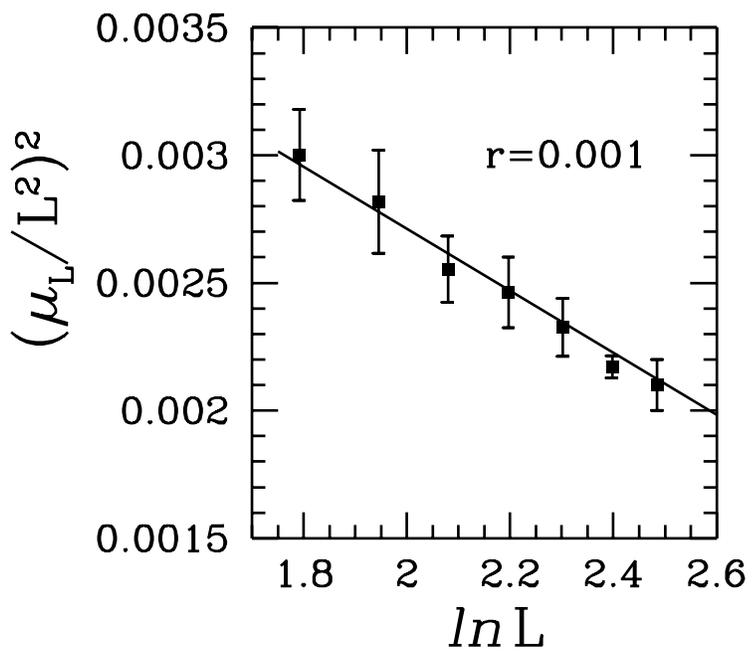
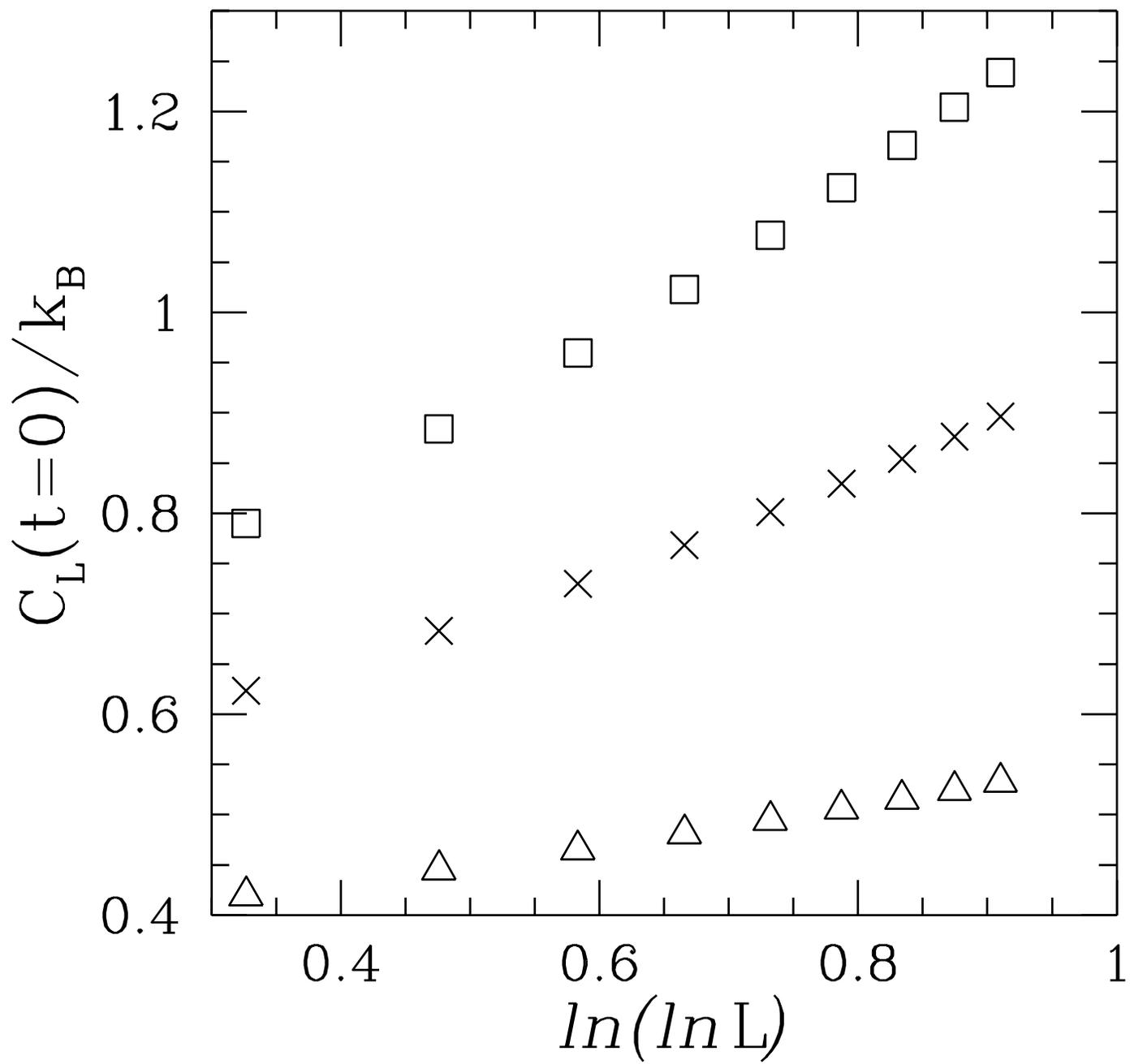
\end{document}